\documentclass[twocolumn]{aastex62}
\usepackage{amsmath}

\usepackage{float}
\graphicspath{{./}{figures/}}

\shorttitle{Transfer of Life Between Earth and Venus}
\shortauthors{Siraj \& Loeb}

\begin{document}
\title{Transfer of Life Between Earth and Venus with Planet-Grazing Asteroids}

\email{amir.siraj@cfa.harvard.edu, aloeb@cfa.harvard.edu}

\author{Amir Siraj}
\affil{Department of Astronomy, Harvard University, 60 Garden Street, Cambridge, MA 02138, USA}

\author{Abraham Loeb}
\affiliation{Department of Astronomy, Harvard University, 60 Garden Street, Cambridge, MA 02138, USA}




\begin{abstract}

Recently, phosphine was discovered in the atmosphere of Venus as a potential biosignature. This raises the question: if Venusian life exists, could it be related to terrestrial life? Based on the known rate of meteoroid impacts on Earth, we show that at least $\sim 6 \times 10^5$ asteroids have grazed Earth's atmosphere without being significantly heated and later impacted Venus, and a similar number have grazed Venus's atmosphere and later impacted the Earth, both within a period of $\sim 10^5$ years during which microbes could survive in space. Although the abundance of terrestrial life in the upper atmosphere is unknown, these planet-grazing shepherds could have potentially been capable of transferring microbial life between the atmospheres of Earth and Venus. As a result, the origin of possible Venusian life may be fundamentally indistinguishable from that of terrestrial life.

\end{abstract}



\section{Introduction}

Recently, the spectral absorption feature of phosphine was discovered in the temperature cloud deck of Venus as a potential biosignature \citep{2020arXiv200906499B, 2020arXiv200906593G, 2020arXiv200906474S, 2020AsBio..20..235S}. If phosphine is produced through biotic, as opposed to abiotic pathways, the discovery could imply a significant biomass in the Venusian atmosphere \citep{2020arXiv200907835L}. This raises the following question: if Venusian life exists, could it be related to terrestrial life?

Panspermia is the conjecture that life can propagate from one planet to another (for a review, see \citealt{2010SSRv..156..239W, 2010IJAsB...9..119W}). Panspermia between the Earth and Venus has been proposed through channels of rock and dust ejecta \citep{1993Metic..28Q.398M, 2005AsBio...5..483G, 2008Ap&SS.317..133W, 2019Ap&SS.364..191J}. The former channel suffers from significant heating that occurs during the passage through the atmosphere, and the latter lacks shielding from harmful radiation during the journey through space. Another version of panspermia alleviates both of these issues by picking up microbes through an atmospheric grazing event \citep{2020IJAsB..19..260S, 2020arXiv200102235S}. The grazing object later impacts another planet, depositing material there as it breaks up. If the biota is deposited when Venus was young and hospitable to life on its surface, then the deposition could be anywhere. Otherwise, seeding events are restricted to the cloud deck at elevation of $50 - 60 \mathrm{\; km}$ \citep{2020arXiv200906474S} where the temperature and pressure are similar to the lower atmospheric conditions on Earth. Such events could have possibly transferred microbes between the atmospheres of Earth and Venus with radiation shielding and without significant heating.

Here, we study the transfer of planet-grazing asteroids between Earth and Venus. In Section \ref{ea}, we investigate the abundance and plausibility of grazing asteroids that were exchanged between Earth and Venus that could possibly transfer life. In Section \ref{ms}, we discuss further details regarding the survival of microbial life during the asteroid's atmospheric passages and journey through space. Finally, in Section \ref{d}, we explore key predictions and implications of our model.

\section{Exchanged Asteroids}
\label{ea}

The recent detection of an Earth-grazing $\sim 60 \mathrm{\; kg}$ asteroid demonstrated that such objects can pass through the atmosphere without significant heating at altitudes of $\gtrsim 85 \mathrm{\; km}$ \citep{2020AJ....159..191S}. As a result, we consider Earth-grazing asteroids that graze through the scale height ($\sim 8 \mathrm{\; km}$) of atmosphere with a lower bound of $\gtrsim 85 \mathrm{\; km}$, due to the significant drop in density beyond a scale height. Asteroids with mass $\sim 60 \mathrm{\; kg}$ strike Earth at a rate of $\sim 1.3 \times 10^3 \; \mathrm{yr^{-1}}$ \citep{2006M&PS...41..607B}, implying that such asteroids pass through a scale height of the Earth's atmosphere at a rate of $\sim 3 \mathrm{\; yr^{-1}}$.

An Earth-grazer typically receives a change of order $\mathrm{1 \; km \; s^{-1}}$ in $v_{\infty}$ through its gravitational interaction with the Earth \citep{2020IJAsB..19..260S}, which is an insignificant change relative to the total orbital energy of near-Earth asteroids. This implies that the fraction of Earth-crossing objects, which are also Venus-crossing\footnote{https://ssd.jpl.nasa.gov/}, remains as observed at $\sim 1/3$ and is largely unaffected by singular gravitational interactions with terrestrial planets. We conservatively assume that the inclinations have changed such that the orbits no longer necessarily lie in the ecliptic plane. The chance that a random Venus-crossing trajectory impacts Venus during one orbit is $(r_V / a_V)^2 \sim 3 \times 10^{-9}$, where $r_V \sim 6 \times 10^3 \mathrm{\; km}$ and $a_V \sim 0.7 \mathrm{\; AU}$ are the radius and semi-major axis of Venus, respectively.

Polyextremophiles such as \textit{Deinococcus radiodurans} are estimated to have an exponential survival probability on a timescale of $\sim 10^5 \mathrm{\; yr}$ with minimal radiation shielding in space \citep{2000Icar..145..391M, 2004IJAsB...3...73B, 2018ApJ...868L..12G}. Given that Venus-crossing asteroids have typical orbital periods\footnote{https://ssd.jpl.nasa.gov/} of $\sim 2 \mathrm{\; yr}$, there are $\sim 5 \times 10^4$ orbits over $\sim 10^5 \mathrm{\; yr}$ during which microbes may survive, resulting in a $\sim 1.5 \times 10^{-4}$ probability of impacting Venus over that timescale. The asteroid belt has been in steady-state for $\sim 3.7 \mathrm{\; Gyr}$ \citep{2015RAA....15..407S}, implying that during this period, $\sim 6 \times 10^5$ Earth-grazing asteroids of mass $\sim 60 \; \mathrm{kg}$ have impacted Venus within $\sim 10^5 \mathrm{\; yr}$ of interacting with the Earth. A similar number of objects grazed a scale height of Venus' atmosphere and later impacted the Earth, since there is only $\sim 5\%$ difference between the radii of Venus and the Earth and the orbital radii are only different by $\sim 30 \%$.

\begin{figure}
  \centering
  \includegraphics[width=1\linewidth]{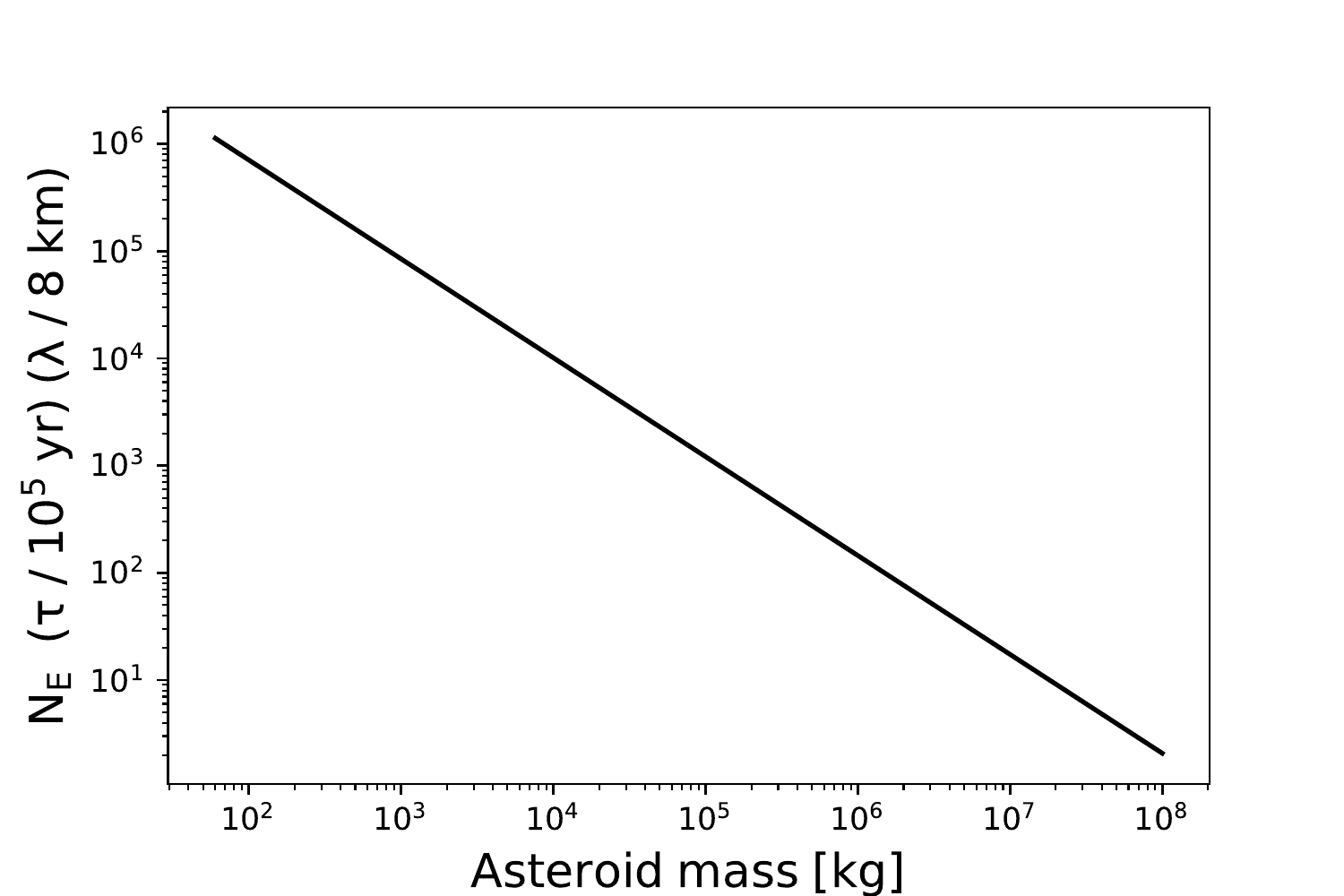}
    \caption{Total number of exchanged planet-grazing asteroids between Earth and Venus, $N_E$, as a function of asteroid mass, per logarithmic bin, scaled to a transfer timescale of $\tau \sim 10^5 \mathrm{\; yr}$ during which microbes may survive and an atmospheric altitude cross section of $\lambda \sim 8 \mathrm{\; km}$, based upon the best-fit of the Earth impact rate presented in \cite{2006M&PS...41..607B}.
}
    \label{fig:exchange}
\end{figure}

The total number of exchanged planet-grazing asteroids as a function of asteroid mass, collision timescale, and atmospheric altitude cross section, is displayed in Figure \ref{fig:exchange}. The number of asteroid impacts on terrestrial planets during Late Heavy Bombardment is not well-understood, however it may have been an order of magnitude higher than the number of impacts while the asteroid belt has been in steady-state \citep{2015RAA....15..407S}, which would increase the number of Earth-grazing and Venus-grazing objects by a similar factor.

\section{Microbial Survival}
\label{ms}

Life has been detected in the Earth's atmosphere up to an altitude of 77 km \citep{Imshenetsky1}. Further work is needed to investigate the existence and abundance of microbial life in the upper atmosphere, particularly at the altitude considered here, $\sim 85 \mathrm{\; km}$, at which Earth-grazing object would avoid significant heating. In addition, if life is discovered by a direct probe sent into the atmosphere of Venus, it will be necessary to calibrate the abundance of life as a function of altitude on Venus.

If accelerated over a distance of $\sim 100 \mathrm{\; m}$, the collected microbes will experience accelerations of order $\sim 10^5 \mathrm{\; g}$. Many terrestrial microbe species have been shown to survive accelerations of $4 - 5 \times 10^5 \mathrm{\; g}$ \citep{2001E&PSL.189....1M, 2011PNAS..108.7997D}, so we assume that acceleration is not an important lethal factor for microbes picked up by the transporting body.

Interiors of objects with radii of $\gtrsim 10 \mathrm{\; cm}$ may not be heated to more than $100 ^{\circ} \mathrm{C}$ during a passage through the Earth's atmosphere \citep{2004MNRAS.348...46N}. This would increase the grazing cross-sections of Earth and Venus that allow for microbial survival, and additionally imply that planet-grazers would not be heated significantly when travelling to the surface of Earth or the cloud-decks of Venus. Many asteroids are known to have significant porosity \citep{2000Icar..146..213B, 2006LPI....37.2214B}, making it possible for microbes to become lodged inside and shielded from exterior heating \citep{2020arXiv200102235S}. While some polyextremophiles can survive in space with minimal radiation shielding, asteroids tens of cm in size may provide sufficient shielding from radiation to protect the survival of a greater diversity microbes \citep{2002abqc.book...57H}. Microbes could be potentially be deposited in clouds when a meteor disintegrates, despite the heating by friction with the atmosphere, due to the effects of shielding by the outer layers of the meteor.

\newpage

\section{Discussion}
\label{d}

We showed that, throughout the history of the solar system, at least $\sim 6 \times 10^5$ asteroids have grazed Earth's atmosphere without undergoing significant heating and later impacted Venus within a microbe survival timescale of $\sim 10^5 \; \mathrm{yr}$, and that a similar number of objects grazed the atmosphere of Venus and later impacted the Earth. If these asteroids picked up microbes during their grazing events, they could have potentially transferred life between Earth and Venus.

Due to the uncertainties regarding the abundance and nature of microbial life in the Earth's atmosphere, in addition to poorly constrained factors relating to microbial survival during pick-up, transport, and delivery, it is not at present possible to determine whether any life was transferred between Earth and Venus. 

A future probe that could sample the habitable cloud deck of Venus will potentially enable the direct discovery of microbial life outside of Earth. Specifically, the capability to either directly analyze microbes \textit{in situ} or to return an atmospheric sample to Earth will be critical in the design of a successful mission. Finding exactly the same genomic material and helicity on Venus and Earth would constitute a smoking gun for panspermia.

This potentially viable mechanism for transferring life between the two planets implies that if Venusian life exists, its origin may be fundamentally indistinguishable from that of terrestrial life, and a second genesis may be impossible to prove. Our model predicts that putative Venusian life would share chemical and biological structures with life on Earth, such as RNA and DNA. Future space missions to Venus will test this hypothesis, and further investigation of the Earth's atmosphere will help refine models of panspermia. Additionally, tests of panspsermia across the rocky planets of the solar system will hold important implications for the likelihood of life on more than one planet per multiplanetary system.

\section*{Acknowledgements}
This work was supported in part by a grant from the Breakthrough Prize Foundation. 





\newpage 

\bibliography{bib}{}
\bibliographystyle{aasjournal}



\end{document}